\documentclass[11pt]{article}

\usepackage{graphicx, verbatim}
\usepackage{amssymb}
\usepackage{alltt}
\usepackage{amsmath, amssymb, amsfonts, amscd, xspace, pifont, amsthm}
\usepackage{mathrsfs}
\usepackage{algorithm}
\newtheorem{theorem}{Theorem}

\newtheorem*{thma}{Theorem A1}

\def\twoImages#1#2#3#4#5#6 
{
  \centerline{\hfill\makebox[#2]{#3}\hfill\makebox[#5]{#6}\hfill}
  \centerline{\hfill
    \includegraphics[width=#2]{#1}
    \hfill
    \includegraphics[width=#5]{#4}
    \hfill}
}

\newcommand{\V}[1]{\ensuremath{\boldsymbol{#1}}\xspace}
\newcommand{\M}[1]{\ensuremath{\boldsymbol{#1}}\xspace}
\newcommand{\F}[1]{\ensuremath{\mathrm{#1}}\xspace}

\title{Community extraction for social networks}
\author{Yunpeng Zhao, Elizaveta Levina, and Ji Zhu \\
Department of Statistics \\
University of Michigan
}
\date{May 17, 2010}

\begin{document}
\maketitle
\begin{abstract}
Analysis of networks and in particular discovering communities within networks has been a focus of recent work in several fields, with applications ranging from citation and friendship networks to food webs and gene regulatory networks.  Most of the existing community detection methods focus on partitioning the entire network into communities, with the expectation of many ties within communities 
and few ties between.  However, many networks contain nodes that do not fit in with any of the communities, and forcing every node into a community can distort results.  Here we propose a new framework that focuses on community extraction instead of partition, extracting one community at a time.  The main idea behind extraction is that the strength of a community should not depend on ties between members of other communities, but only on ties within that community and its ties to the outside world.  We show that the new extraction criterion performs well on simulated and real networks, and establish asymptotic consistency of our method under the block model assumption.

\end{abstract}
\section{Introduction}
Understanding and modeling network structures has been a focus of attention in a number of diverse fields, including physics, biology, computer science, statistics, and social sciences.  Applications of network analysis include friendship and social networks, marketing and recommender systems, the world wide web, disease models, and food webs, among others.  A fundamental problem in the study of networks is community detection (see, for example, \cite{NewmanPNAS} for a comprehensive recent review).  The extensive literature on the subject typically assumes that networks consist of communities, which are thought of as tightly-knit groups with many connections between the group members and relatively few connections between groups.  We focus here on 
undirected networks $N=(V,E)$, where $V$ is the set of nodes and $E$ is the set of edges, possibly weighted.   The community detection problem is typically formulated as finding a partition $V = V_1 \cup \dots \cup V_K$ which gives ``tight'' communities in some suitable sense (several examples will be discussed below). 
The node sets $V_1, \dots, V_K$ are typically taken to be disjoint, although there is some recent work on detecting overlapping communities \cite{Palla2005, Adamcsek2006, Wei2009}.  Whether communities are overlapping or not, every node is required to belong to at least one community.  There are many examples of networks where such a requirement makes sense, for example, the college football games network \cite{Girvan&Newman2002}, and yet some commonly studied networks clearly do not fit this framework.  For example, in the high school friendship network of \cite{Hunter2008} discussed later in the paper, there are people who belong to tight communities, but there are also people who do not have ties to any community at all.  There is surprisingly little work allowing for such a network structure.

In this paper, we propose broadening the framework of community detection to allow for a network to contain not only communities, but also ``background'' nodes that are not required to have tight connections to anything else in the network.   We approach the problem via sequential community extraction rather than network partition: at each step, we extract the tightest (in a certain sense to be defined) community with the sparsest connections to the rest of the network.  The process is repeated for the remaining nodes until no more meaningful communities can be extracted; the remainder of the network is then classified as background.  Community extraction can also be used in conjunction with partition, for example, to identify the cores of communities found by a partition algorithm.  
The key characteristic of our approach is that we focus on edges within the 
candidate community to be extracted and edges connecting it to the rest of the network, and ignore edges within the rest of the network. The intuition here is that edges not related to nodes in a potential community should not influence our judgment on the community in question.   A by-product of this process is a ranking of extracted communities, which standard partitioning methods do not provide.  Extraction may also be more robust to changes in the network over time, 
since the definition of a community does not rely on links between unrelated nodes. 

\begin{figure}[h!]
\begin{center}
\twoImages{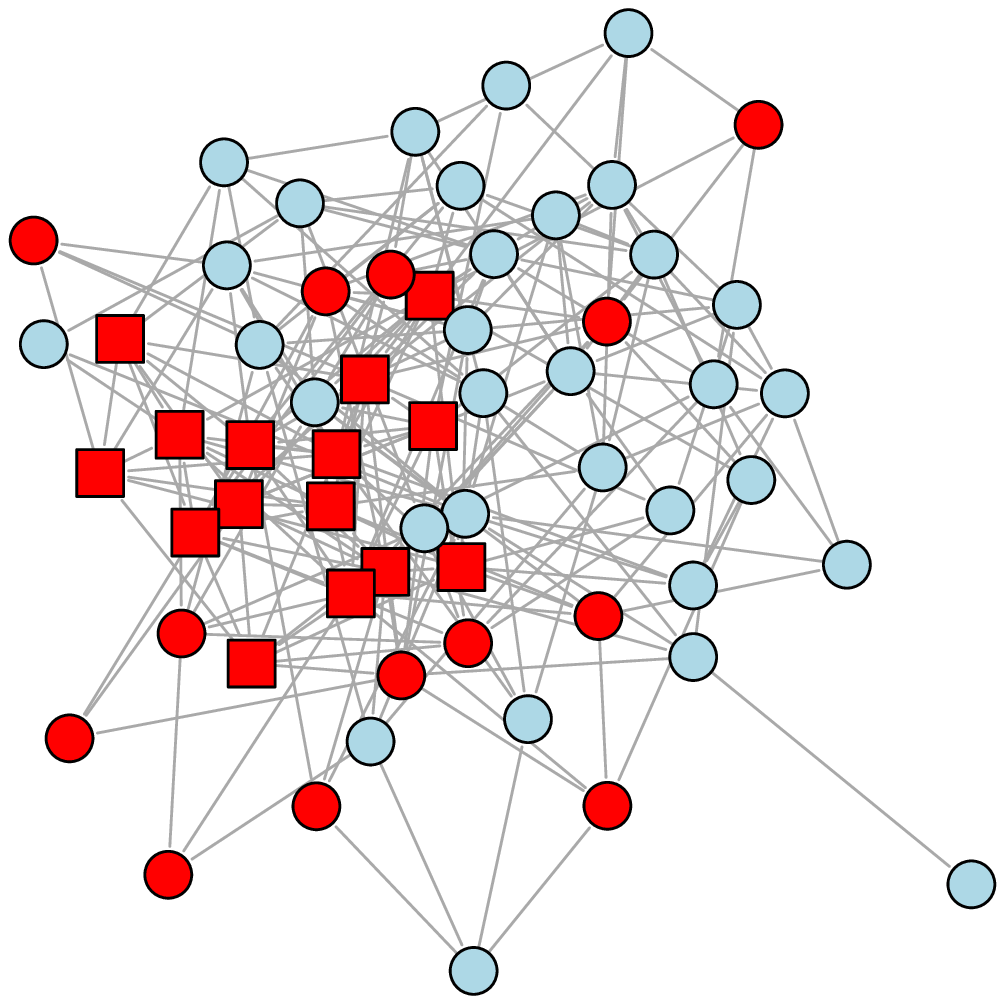}{5cm}{(a) Partition}
{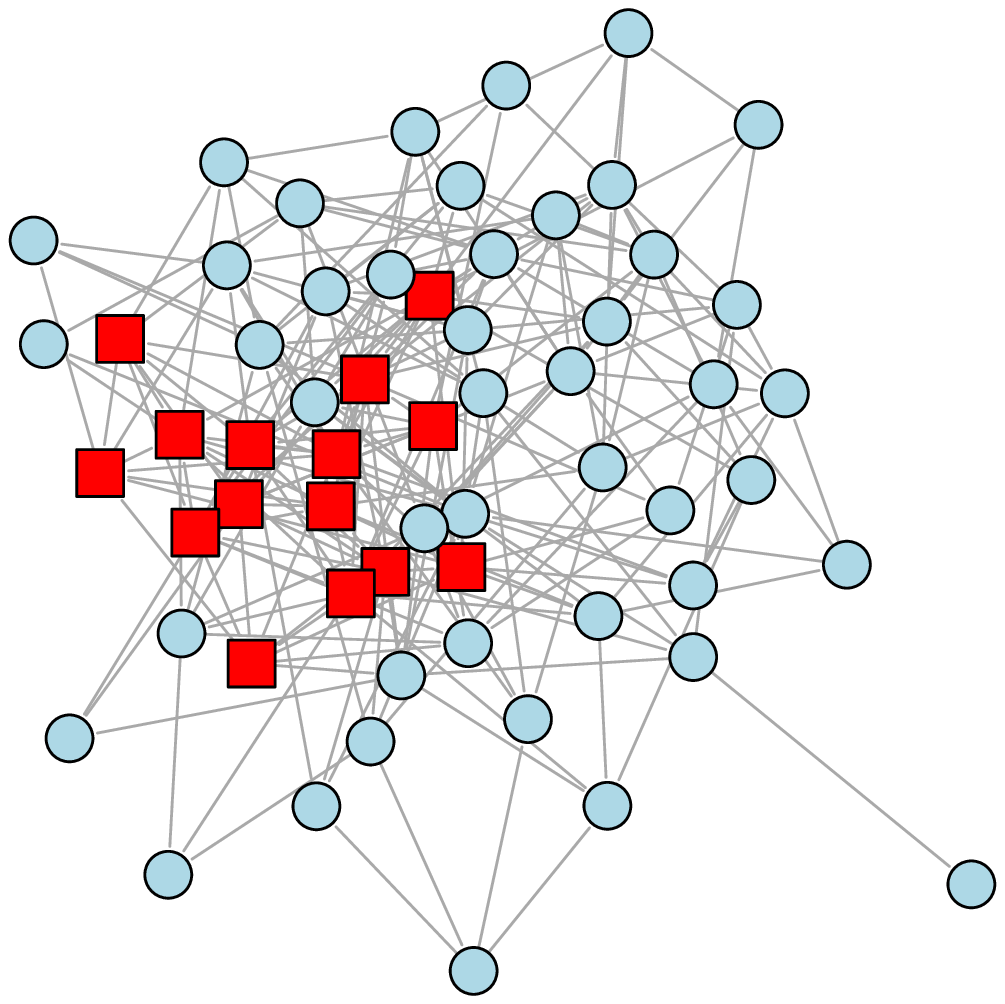}{5cm}{(b) Extraction}
\caption{Toy example: shapes represent the truth and colors represent 
partition results using modularity (left) and extraction results using our method (right).}
\label{fig:toy}
\end{center}
\end{figure}

To illustrate the motivation for our approach, consider the following toy example: out of $n = 60$ nodes, 15 belong to a community where links between members form independently with probability 0.5 each.  The links from members to the other 45 nodes and links between the other 45 nodes all form independently with probability 0.1.  Results of partition (using the modularity method of \cite{Newman&Girvan2004}) into two communities and community extraction by our method are shown in Figure \ref{fig:toy}.  Partitioning has to balance tightness of the two communities, and as a result includes a number of background nodes in the community.  Extraction, on the other hand, separates the community out perfectly.  

Most popular community detection methods focus on maximizing links within communities while minimizing links between communities.  This can be achieved either implicitly through an algorithmic approach (see 
\cite{Newman2004Review} for a review) or explicitly by optimizing a criterion that measures quality of a proposed partition over all possible partitions.  The criteria proposed include ratio cuts \cite{Wei&Cheng1989}, normalized cuts \cite{Shi&Malik2000}, spectral clustering \cite{Ng01}, and modularity \cite{Newman&Girvan2004, NewmanPNAS}.  All of
these methods are designed to partition networks that consist of pure communities, 
with no background.   Another class of methods assumes a parametric statistical model for generating the network and estimates the partitioning by maximizing the likelihood or by employing a Bayesian method.  The models used for partitioning include the block model \cite{Snijders&Nowicki2000, Bickel&Chen2009}, a mixture model 
\cite{Newman&Leicht2007}, univariate \cite{Hoff2002} and multivariate \cite{Handcock2007} latent variable models; for a comprehensive review of statistical models of networks, see \cite{Goldenberg2010}.   While each statistical model has its advantages and often aids in interpretation of the data, it also imposes its own constraints and assumptions, in contrast to graph partitioning methods that do not typically depend on model assumptions (although in practice they 
may work well for some models and fail for others).   
Our method is also based on a model-free non-parametric criterion, and we will show  that it is consistent under one of the commonly used statistical models, the stochastic block model.  Finally, another line of work 
related to ours is the 
core-periphery partition methods \cite{Borgatti&Everett1999,Everett&Borgatti1999}.  These authors 
use a different criterion to separate a tight ``core'' from a sparse ``periphery'', whereas our criterion is designed to extract the tightest community regardless of whether the rest of the network is sparse or contains other communities.  

In the remainder of the paper, we present the new methodology for community extraction, show that it is asymptotically consistent  under the stochastic block model, and apply the method to a number of simulated and real networks, comparing extraction and partition results.  The proofs of theorems are given in the Supplementary Material.

\section{The community extraction methodology}

\label{sec:meth}

To focus ideas, we start from discussing several related partitioning methods.   
An undirected network $N=(V,E)$ with $|V| = n$ nodes can be represented by an $n \times n$ adjacency matrix $\M{A} = [A_{ij}]$, where $A_{ij} > 0$ if there is an edge between nodes $i$ and $j$ and $A_{ij} = 0$ otherwise.  If the network has weights associated with edges, the positive $A_{ij}$'s are the weights; if not, the positive $A_{ij}$'s are set to 1.  Since we focus on undirected networks, $\M{A}$ is symmetric.  
For simplicity, we focus on partition into two sets $(V_1,V_2)$, where $V_1 \cap V_2 = \emptyset$ 
and $V = V_1 \cup V_2$.  A partition is associated with a partition 
vector $\V{s}$, where $s_i= 1$ if node $i$ belongs to $V_1$, and 
$s_i = -1$ if node $i$ belongs to $V_2$.
              
A naive way to partition a network is to minimize the total weight $R$ of edges connecting $V_1$ and $V_2$ over all possible partitions (the min-cut method).  The total weight, or the cut, is given by 
\begin{equation}
R= \sum_{i \in V_1, j \in V_2} A_{ij} \ . 
\end{equation}
However, minimizing $R$ yields a trivial solution of $V_1 = V$.  
The ratio cut approach \cite{Wei&Cheng1989} avoids the trivial solution by 
minimizing 
 $R/(|V_1|\cdot |V_2|)$, where $|V_1|$ and $|V_2|$ are the sizes of the two groups. Efficient spectral algorithms for computing the ratio cut are available \cite{Hagen&Kahng1992}.  Another approach is minimizing the  normalized cut \cite{Shi&Malik2000}, initially proposed for image segmentation.  The normalized cut is defined as  $R/D_1 + R/D_2$, where $D_k=\sum_{i \in V_k, j \in V} A_{ij}$ for $k = 1,2$ is the total number of edges involving nodes in $V_k$.  Dividing by $D_k$ encourages balanced group sizes and avoids trivial solutions. The normalized cut criterion can be approximated by a generalized eigenvalue problem and thus solved efficiently.

In the context of community detection in networks, perhaps the most popular criterion for partitioning  is modularity \cite{Newman&Girvan2004}. The intuition behind the modularity criterion is to compare the observed number of edges within groups to the expected number under the configuration model of  \cite{Chung&Lu2002}.  Under this model, an edge between nodes $i$ and $j$ is created independently of other edges with probability $P_{ij} = k_i k_j / 2m$, where 
where $k_i = \sum_j A_{ij}$ is the degree of node $i$ and $2m = \sum_i k_i$ is twice the number of edges in the network.  The modularity criterion is then defined as 
\begin{equation}
Q= \frac{1}{4m} \sum_{ij} [A_{ij} - P_{ij}] s_i s_j \ ,
\end{equation}
where the multiplier $1/4m$ is a matter of convention. Like the normalized cut, the modularity solution can be approximated using the eigen-decomposition of the corresponding modularity matrix $A_{ij} - P_{ij}$ 
\cite{Newman2006}, and thus computed efficiently. In fact, in practice the normalized cut and the modularity solutions tend to be very similar, even though they are motivated by quite different considerations.

For simplicity, we have given all the criteria above for partitioning into two communities; the normalized cut and modularity can easily be adjusted to partition into $K$ communities.   Typically, however, $K$ is unknown and estimating $K$ is an open problem.  Another option, more common in practice, is to proceed sequentially: split the network into two groups, then split each group further, and so on.  This greedy method may miss the optimal partition.  It also requires a stopping criterion; for modularity, a natural rule is to stop when a proposed split decreases the overall modularity \cite{Newman2006}.

Regardless of the criterion used, partitioning methods are not designed to deal with the situation when background nodes are present, without tight links to any part of the network.  Typically, such nodes will be split and grouped together with tighter communities present in the network, rather than separated out in a class of their own.  This is mainly because partitioning methods are symmetric -- the sets $V_1$ and $V_2$ can be interchanged in any of the above criteria.  However, if the goal is to separate a tight community from a sparse background, the roles of $V_1$ and $V_2$ cannot be the same -- in fact, they have to be the opposite.  This observation lies at the core of our proposed methodology.  

\subsection{The community extraction criterion}

The criterion we propose extracts one community at a time
by looking for a set with a large number of links within itself and a small number of links to the rest of the network.  The links within the complement of this set do not matter, and thus the remainder can contain background nodes and/or other communities.  To emphasize the lack of symmetry in the criterion, we 
denote the community to be extracted by $S$ and its complement by $S^c$ 
(rather than $V_1$ and $V_2$).  Then we maximize the following  
extraction criterion over all possible $S$:
\begin{equation}\label{original}
W(S) =   \frac{O(S)}{|S|^2} - \frac{B(S)}{|S||S^c|} \ ,
\end{equation}
where 
\begin{align*}
O(S)  = \sum_{i,j\in S} A_{ij} \ , \  B(S) = \sum_{i\in S,j \in S^c} A_{ij} \ .
\end{align*}  
The term $O(S)$ is twice the number of the edges within $S$, and $B(S)$ counts 
connections between $S$ and the rest of the network. Each term 
is normalized by the total number of possible edges in each case, which gives 
these quantities a natural interpretation as probability estimates under the stochastic block model, discussed further below. Note that we ignore the issue of self-loops and normalize the first term by $|S|^2$ rather than $|S|(|S|-1)$; in practice this makes little difference.

One drawback of criterion \eqref{original} is that, like the original graph cut, it does not explicitly guard against splitting off small communities.  The trivial solution does not maximize $W$, but nonetheless, in a large sparse network a very small community can give a high value of $W$, since the second term will be made negligible by the large $|S^c|$ in the denominator.   To avoid this situation, we can use an adjustment in the spirit of the ratio cut, and maximize the following 
criterion instead: 
\begin{equation}\label{modified}
W_a(S) =  |S||S^c| \left[ \frac{O(S)}{|S|^2}-\frac{B(S)}{|S||S^c|} \right ]  \ . 
\end{equation}
Since $|S||S^c|$ is maximized at $|S|=n/2$, this factor penalizes very small and very large communities and produces more balanced solutions. 
Empirically, we found that the adjustment makes a difference in sparse 
networks, but plays no role in dense networks.  Later we show that asymptotically both criteria are consistent.

Our community extraction procedure consists of sequentially applying criterion \eqref{modified}: we  extract a community and apply the extraction again to its complement.  If there is prior information on the correct or desired number of communities to be extracted, we stop after this number has been obtained and declare the rest to be background.  In the absence of such information, we continue  until we cannot find a community bigger than a certain preset size (for the examples in this paper, 5 nodes), and classify the rest as background. 
This procedure is greedy, but it differs from the usual 
sequential graph partitioning in that we never split a community further 
once it has been extracted, and thus we are not in danger of producing a 
solution consisting of a large number of small communities.

\subsection{Maximizing the extraction criterion}

Finding the exact global maximum of the extraction criterion is NP-hard.  Here we use a local optimization technique based on label switching known as tabu search \cite{Beasley1998, Glover&Laguna1997}. 
The key idea of tabu search is that once a node label has been switched, it cannot be switched again for the next $T$ iterations (the node has ``tabu'' status). This guards against being trapped in a local maximum.  The algorithm starts from an initial value and examines all current non-tabu 
nodes in order.  If the current value of the global maximum can be improved, the node label is switched, its status changed to tabu, and the algorithm returns to node 1.  If no node can be switched to improve the global 
maximum,  the node that gives the largest increase 
in the current criterion value is switched, and if no increase is possible, the node that gives the smallest decrease is switched. The algorithm is run for a prescribed number of iterations, and the best solution seen in the course of these iterations is taken to be the final solution (not necessarily the one from the last iteration).   Note that the value of the criterion \eqref{modified} can be updated efficiently in $O(n)$ operations for a single label switch. Finally, since the algorithm depends on the order of nodes as well as on the initial value, we run it for a number of random starting values and random orders of 
nodes.

\section{Asymptotic Consistency}
\label{sec:theory}

Our algorithm does not explicitly rely on a stochastic model for the network, but if we assume the network has been generated by a block model \cite{Holland83, Snijders&Nowicki2000}, we can establish asymptotic consistency of the extraction criterion using the recent results of \cite{Bickel&Chen2009}. The general block model assumes that nodes belong to one of $K$ blocks, and the network is generated as follows.   First, each node is assigned to a block independently of other nodes, with $P(c_i = k) = \pi_k$,  $1\leq k \leq K$, $\sum_{k=1}^K \pi_k=1$, 
where $\V{c} = (c_1, \dots, c_n)$ is the $n\times 1$ vector of labels representing node assignments to blocks.  Then, conditional on $\V{c}$, edges are generated independently with probabilities $P[A_{ij} = 1|c_i=a, c_j=b] = p_{ab}$.    
The vector of probabilities $\V{\pi} = \{\pi_1, \dots, \pi_K\}$ and the  $K \times K$ symmetric matrix $\M{P} = [p_{ab}]_{1\leq a,b\leq K}$  together specify a block model.  We can stipulate the presence of background by requiring, for instance, that $p_{aK} < p_{bb}$ for all $a = 1, \dots, K$, all $b = 1, \dots, K-1$. One could further assume that $p_{aK} = p$ for all $a = 1, \dots, K$, but this assumption is not necessary for the theory to be developed.  

We consider asymptotic consistency of label assignments by the extraction method as the number of nodes $n \rightarrow \infty$.  If $\M{P}$ does not change with $n$, the network 
will become very dense as $n$ grows, so we allow $\M{P}_n$ to depend on $n$.    A natural parametrization is $\M{P}_n = \rho_n \M{P}$, where $\rho_n  
= P[A_{ij}=1] \rightarrow 0$ is the probability of an edge between arbitrary nodes $i$ and $j$.   The expected node degree $\lambda_n = n \rho_n$ becomes the natural parameter to control as $n \rightarrow \infty$.  

Bickel and Chen \cite{Bickel&Chen2009} developed a general framework for checking whether a community-finding criterion can correctly recover the true node labels as $n \rightarrow \infty$, under the block model assumption. 
The details of their conditions for asymptotic consistency are given 
in the Supplementary Materials.  Briefly, the main condition is 
that the proposed criterion is maximized by the true label assignment when all the sample quantities in the criterion are replaced by their population equivalents, which can be viewed as 
a special case of the general theory of minimum contrast estimation \cite{Bickel&Doksum}.  

Since we perform community extraction sequentially, we focus on checking consistency for the case $K=2$ (one extracted community plus the rest of the network).  The matrix $\M{P}$ is $2 \times 2$ with three unique parameters $p_{11}$, $p_{22}$, $p_{12}$,  and the vector of class probabilities $\{\pi, 1 - \pi\}$ is determined by the single parameter $\pi$.    Theorem \ref{Thm for original} gives consistency of criterion \eqref{original}.
\begin{theorem}
\label{Thm for original}
Suppose $\frac{\lambda_n}{\log{n}} \rightarrow \infty$, then for any $0<\pi<1$, if $p_{11}>p_{12}$, $p_{11} > p_{22}$ and $p_{11}+p_{22}>2p_{12}$, the maximizer $\hat{\V{c}}^{(n)}$ of criterion \eqref{original} satisfies  
\begin{align*}
P[\hat{\V{c}}^{(n)}=\V{c}] \rightarrow 1 \quad as \quad  n \rightarrow \infty,
\end{align*} 
where $\V{c}$ are the true labels. 
\end{theorem} 
Note that the simplest case of one community with background nodes connecting at random to all nodes in the network ($p_{12}=p_{22} = p$) is covered by the theorem as long as $p_{11} > p$.

The adjusted criterion \eqref{modified} differs from the original criterion \eqref{original} by a factor of $|S||S^c|$, but it turns out that this factor has no effect in the limit, and asymptotic consistency holds for the adjusted criterion as well. 
\begin{theorem}\label{Thm for adjusted}
\emph{(Adjusted criterion)}
Suppose $\frac{\lambda_n}{\log{n}} \rightarrow \infty$, for any $0<\pi<1$, if $p_{11}>p_{12}$, $p_{11} > p_{22}$ and $p_{11}+p_{22}>2p_{12}$, the maximizer $\hat{\V{c}}^{(n)}$ of criterion \eqref{modified} satisfies  
\begin{align*}
P[\hat{\V{c}}^{(n)}=\V{c}] \rightarrow 1 \quad as \quad  n \rightarrow \infty.
\end{align*} 
\end{theorem} 
Proofs of both theorems are given in the Supplementary Materials.

\section{Numerical evaluation}
\label{sec:sim}

In this section, we compare performance of our original extraction 
criterion, adjusted extraction criterion and modularity for three different simulated scenarios. 
We compare the methods using the positive predictive value (PPV) and 
the negative predictive value (NPV), defined as follows. 
Let $S$ be the extracted community, and let $C_S$ be the true community that 
matches $S$ best, determined by majority vote.  
Then we define 
\begin{align*}
\mbox{PPV} = & \frac{|C_S\cap S|}{|S|} \\
\mbox{NPV} = & 1-\frac{|C_S\cap S^c|}{|S^c|}
\end{align*}
The PPV is a measure of purity of the extracted community, and the NPV is a measure of completeness.

First we consider the case of two communities with no background, to check that our method works in this standard situation. Specifically, we generate  networks with 1000 nodes from a block model with $K=2$, $p_{11} = 0.5$, $p_{22} = 0.4$, and $p_{12}= 0.05$, with fixed first community size $n_1$. 
The number of replications in this simulation and all following is fixed at 50.
Table \ref{tab:twoclusters} shows the means and standard deviations of 
PPV and NPV over 50 replications for $n_1 = 100$; for more balanced community sizes ($n_1 = 200$ and larger) all methods find the ideal partition, and these results are not shown.   

\begin{table}[h!]
\begin{center}
\caption{Results for two communities with no background: mean(SD) of positive and negative predictive values over 50 replications. }
\label{tab:twoclusters}
\begin{tabular}{c|ccc}
   & Modularity & Original & Adjusted \\
\hline
PPV & 1(0)       & 1(0) & 1(0) \\
NPV & 0.84(0.03) & 1.00(0.00) & 0.71(0.06) \\
\end{tabular}
\end{center}
\end{table}

All methods do perfectly in terms of PPV, meaning all nodes in the extracted community belong to the same true class.  The original criterion gives the best performance on NPV (meaning no nodes from the true community were ``lost'' in extraction),
since this network is dense, and the problem with splitting off small clusters does not occur.  The adjusted criterion performs worse (although for larger $n_1$ all methods perform the same); however, the opposite phenomenon occurs for 
relatively sparse networks, which is illustrated in the next example.   

In the second simulation, we consider sparse networks with one community 
and background, generated from the block model.  Again we fix the number of nodes at 1000, and let $p_{12} = p_{22}= 0.05$.   We consider three community sizes ($n_1 = 100$, 200, 300), and three values of $p_{11} = 0.1$, 0.15, and 
0.2.   
In the third simulation, we consider a more complicated situation with two communities with similar densities and a sparse background.  The network size is fixed at 1000 and the probability that a background node has a link to  any other node is 0.05.  We consider two different sizes for both communities 
(100 and 300), and three values of $p_{11}=0.05 x$, $p_{22}=0.04 x$, for $x=2$, 3, 4.   Here we compare the results of extracting one community with partitioning into two communities using modularity.

\begin{figure}[h!]
\begin{center}
 \includegraphics[width=8.0cm]{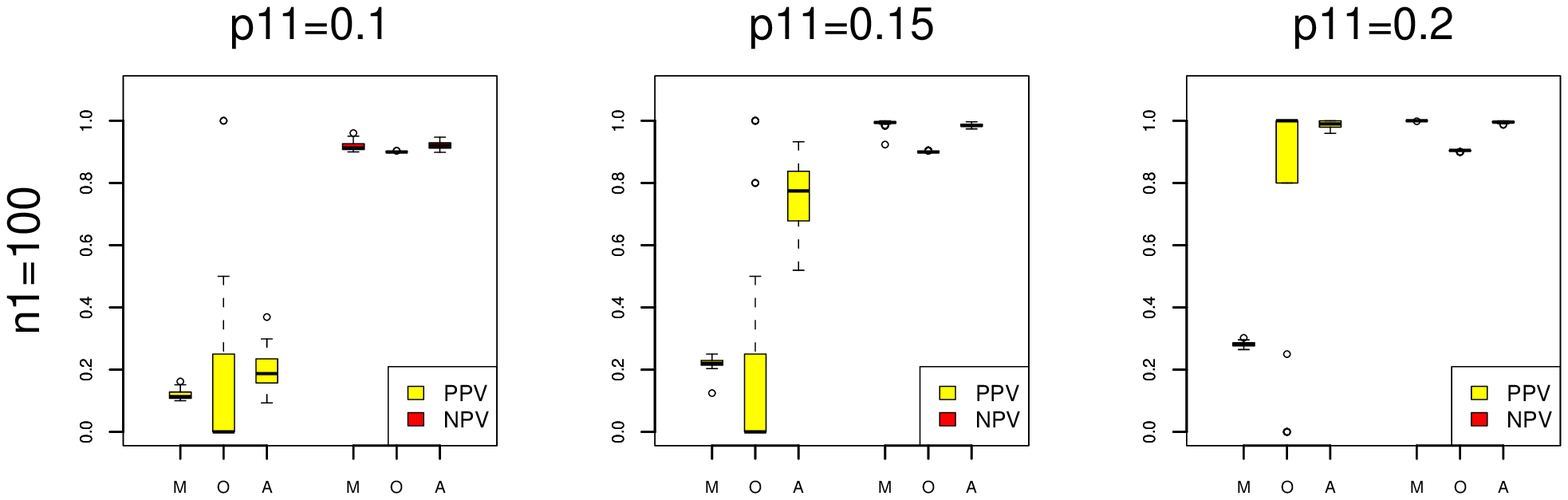}\\
 \includegraphics[width=8.0cm]{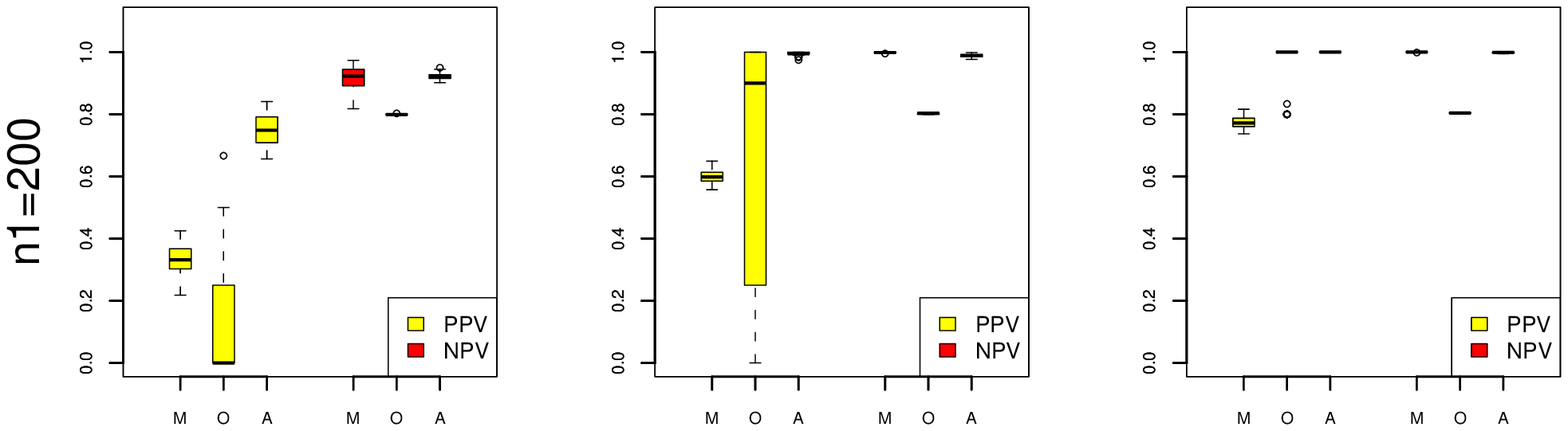}\\ 
 \includegraphics[width=8.0cm]{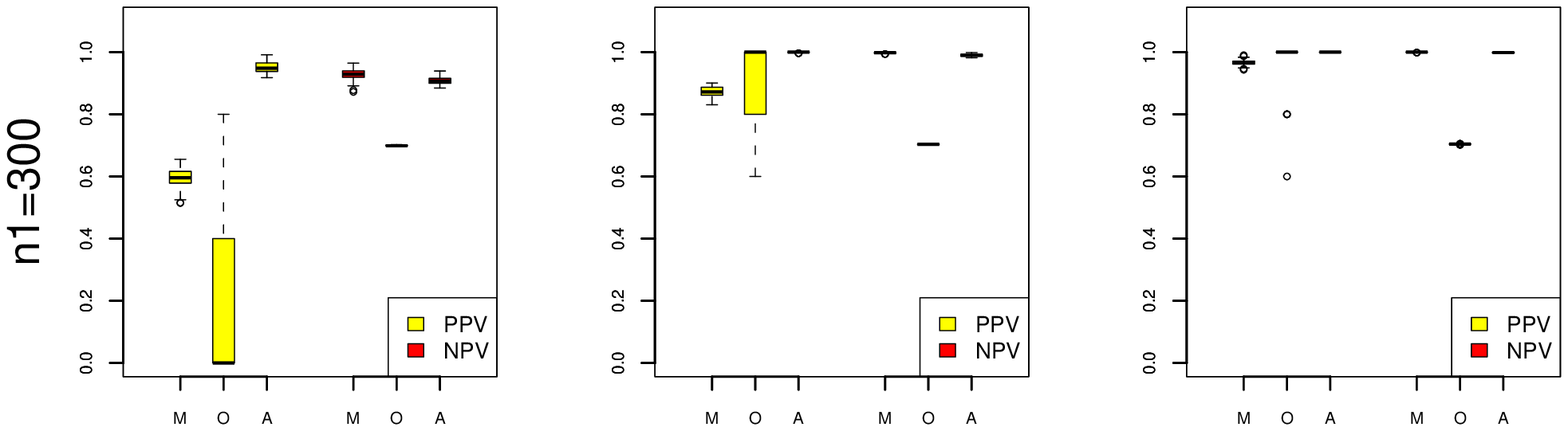}
\caption{Boxplots of PPV and NPV for one community with background. M: Modularity; O: Original extraction criterion; A: Adjusted extraction criterion}
\label{fig:onecluster+noise}
\end{center}
\end{figure}

\begin{figure}[h!]
\begin{center}
 \includegraphics[width=8.0cm]{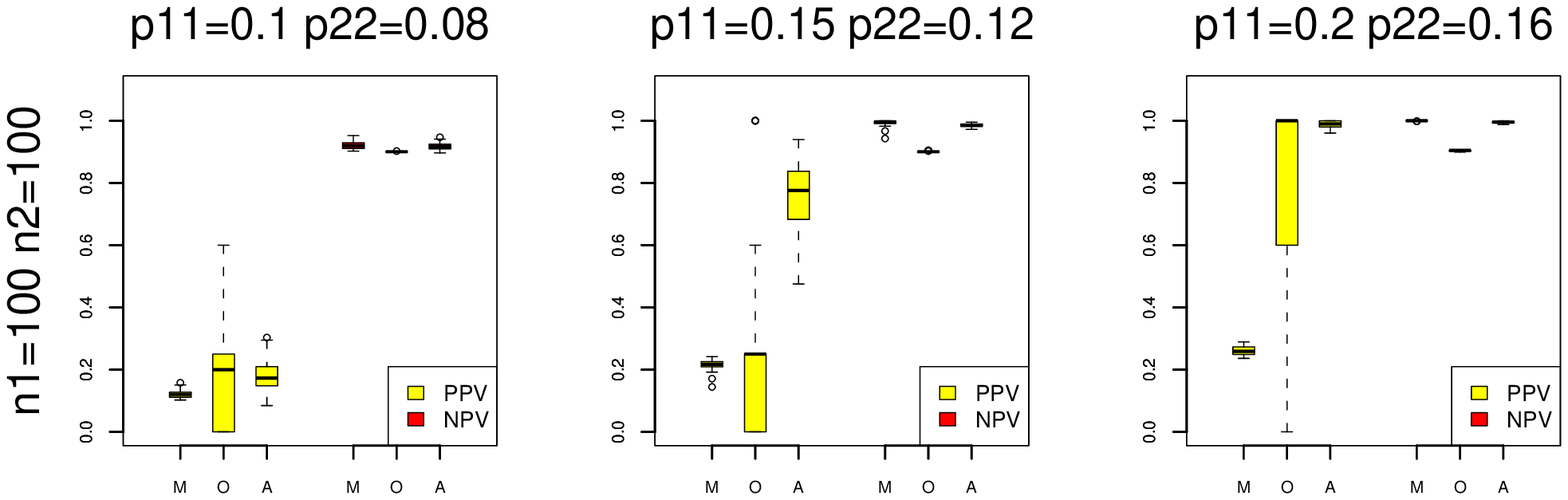}\\
 \includegraphics[width=8.0cm]{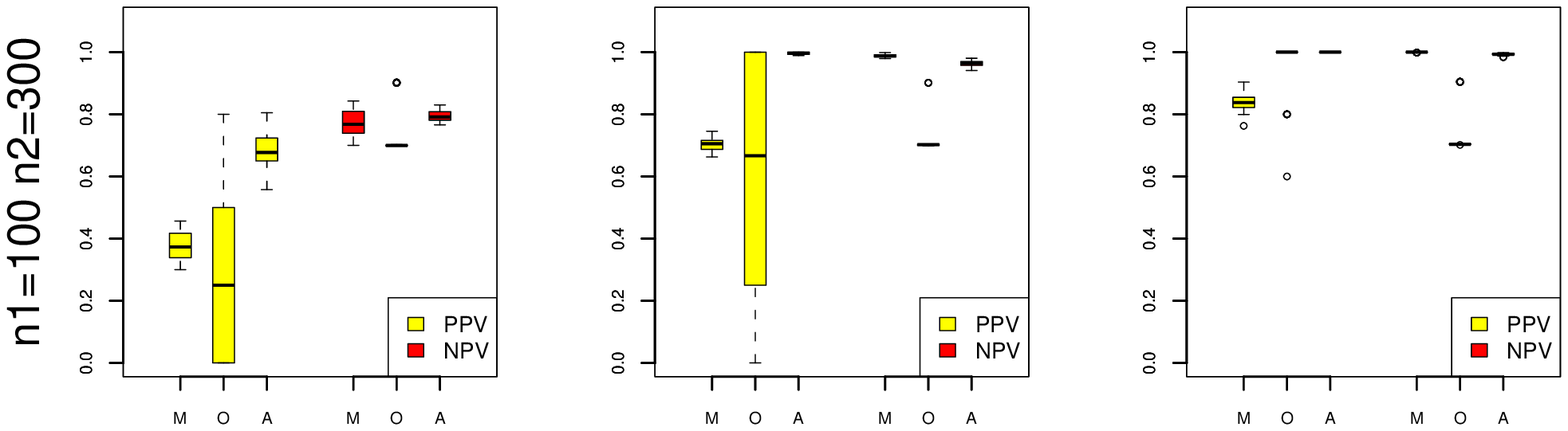}\\ 
 \includegraphics[width=8.0cm]{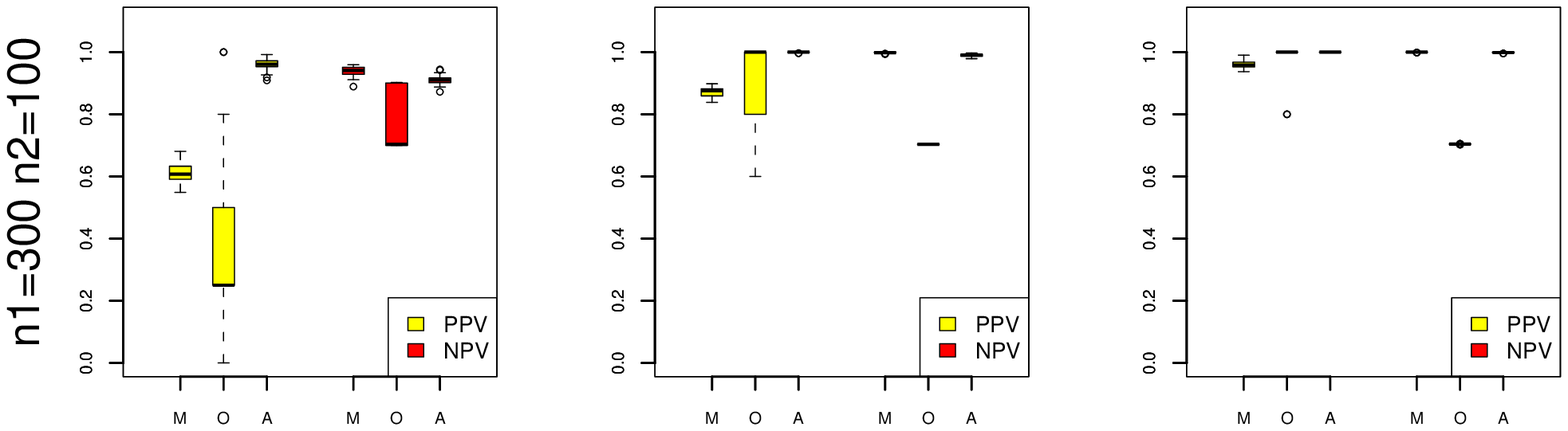}\\
 \includegraphics[width=8.0cm]{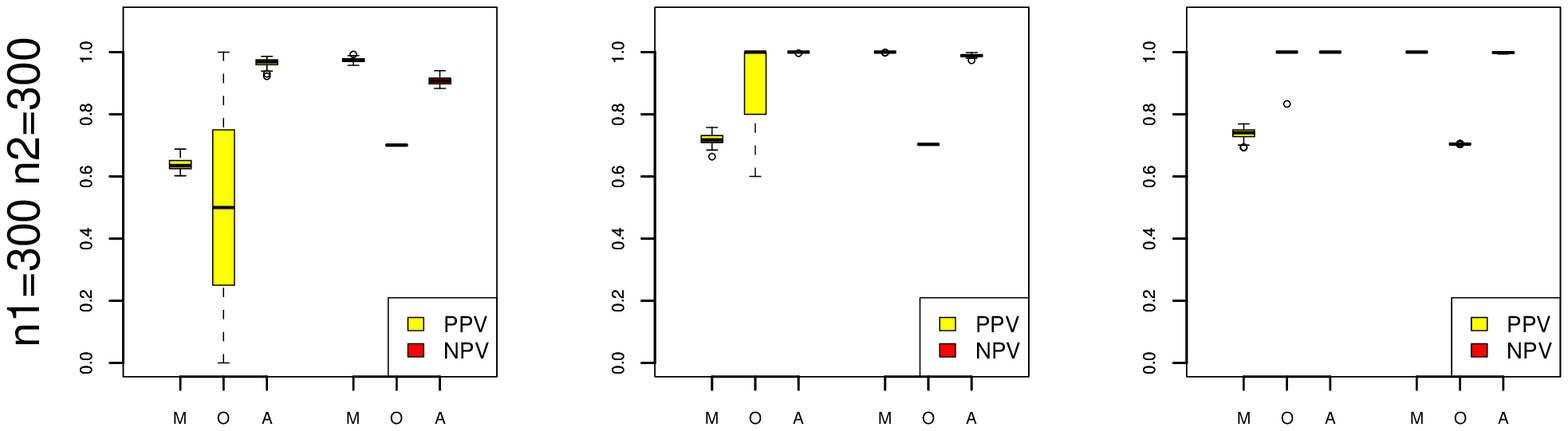}
\caption{Boxplots of PPVs and NPVs for two communities with background. M: Modularity; O: Original extraction criterion; A: Adjusted extraction criterion}
\label{fig:twoclusters+noise}
\end{center}
\end{figure}

The results for the second and third simulations are presented in Figures \ref{fig:onecluster+noise} and \ref{fig:twoclusters+noise}, respectively, 
with boxplots of the 50 replication values of NPV and PPV shown side by side for all methods. For the case of weakly connected communities ($p_{11}=0.1$), the original 
criterion tends to favor small clusters, which is evident from the NPV values.
On the other hand, the adjusted criterion performs best overall  -- even though the NPVs of the adjusted criterion are slightly lower than those of modularity, the PPVs are significantly higher, particularly for weak signals 
(small $n_1$ and low $p_{11}$).  This is because the background nodes are roughly equally split between the two parts found by modularity. 

\section{Examples}
\label{sec:data}

In this section, we apply our extraction method (using the adjusted criterion) 
to several real-world networks and compare results with partition as computed by modularity.  We perform 
extraction sequentially, and each time use ten different starting 
values for the tabu search. We stop the extracting procedure when the 
proposed community has fewer than five nodes.  

\subsection{The karate club network}
\begin{figure}[h!]
\begin{center}
\twoImages {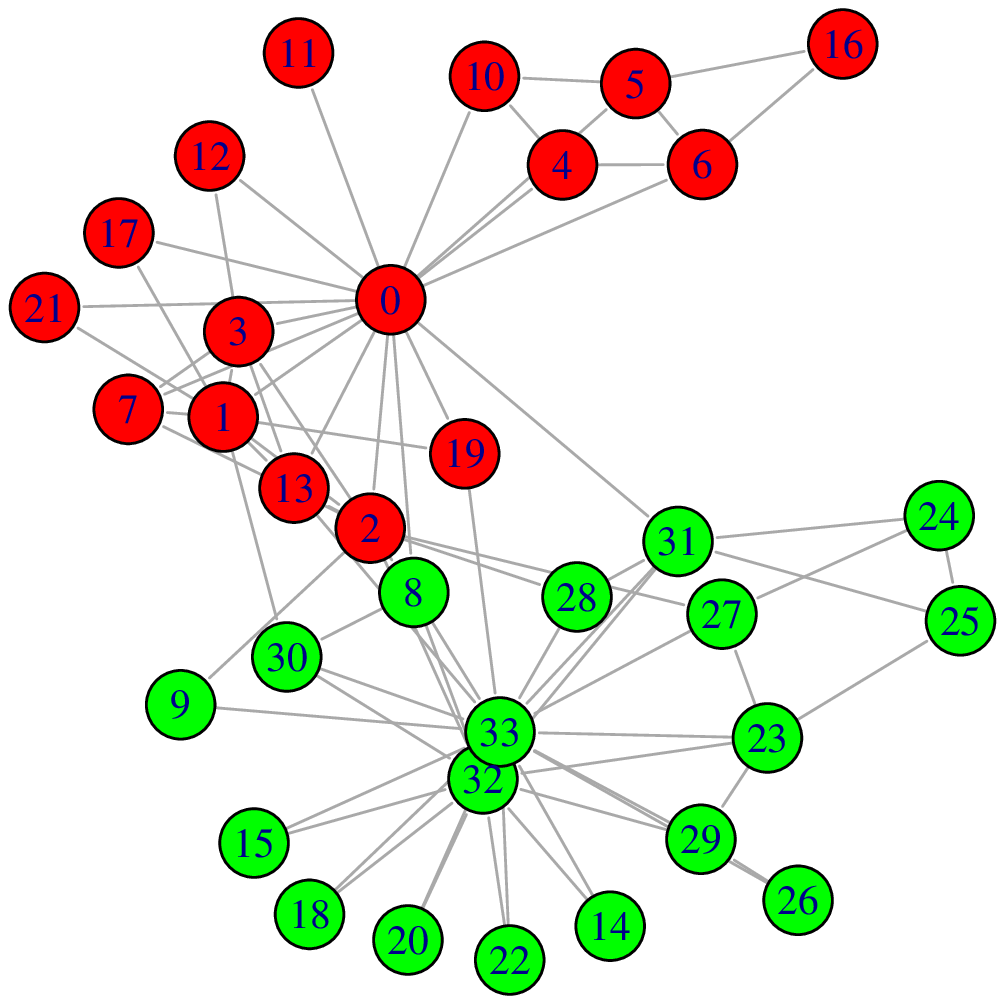}{5cm}{(a) Partition}
           {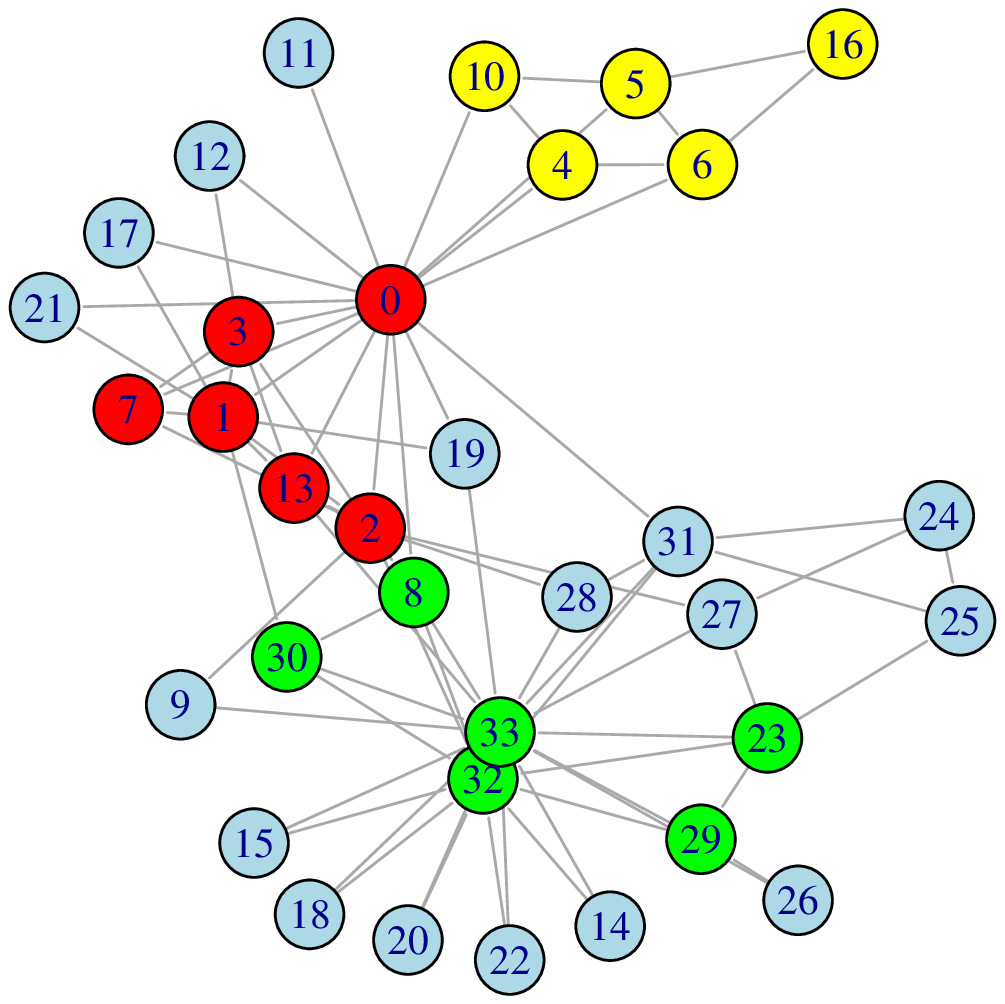}{5cm}{(b) Extraction}
\caption{Results for the karate club network}\label{fig:karate}
\end{center}
\end{figure}

Our first example is a well-known friendship network representing friendships between 34 members of a karate club \cite{Zachary1977}.  This club had 
subsequently split into two parts following a disagreement between an 
instructor (node 0) and an administrator (node 33), and these two groups 
are used as the ``ground truth'' in benchmark studies of community detection algorithms.  Modularity partitions this network into exactly the true factions \cite{NewmanPNAS}. The extraction approach can be used to supplement this division with more information by identifying the ``cores'' of each faction. 
Extraction found three groups --  
the cores of two factions, which contain the instructor and the administrator (shown in green and red in Figure \ref{fig:karate}), 
and a small tight community within the instructor's faction (shown in yellow in Figure \ref{fig:karate}).  Note that there are no extracted communities 
that mix the members of the two factions.  

\subsection{The political books network}

\begin{figure}[h!]
\begin{center}
\twoImages
{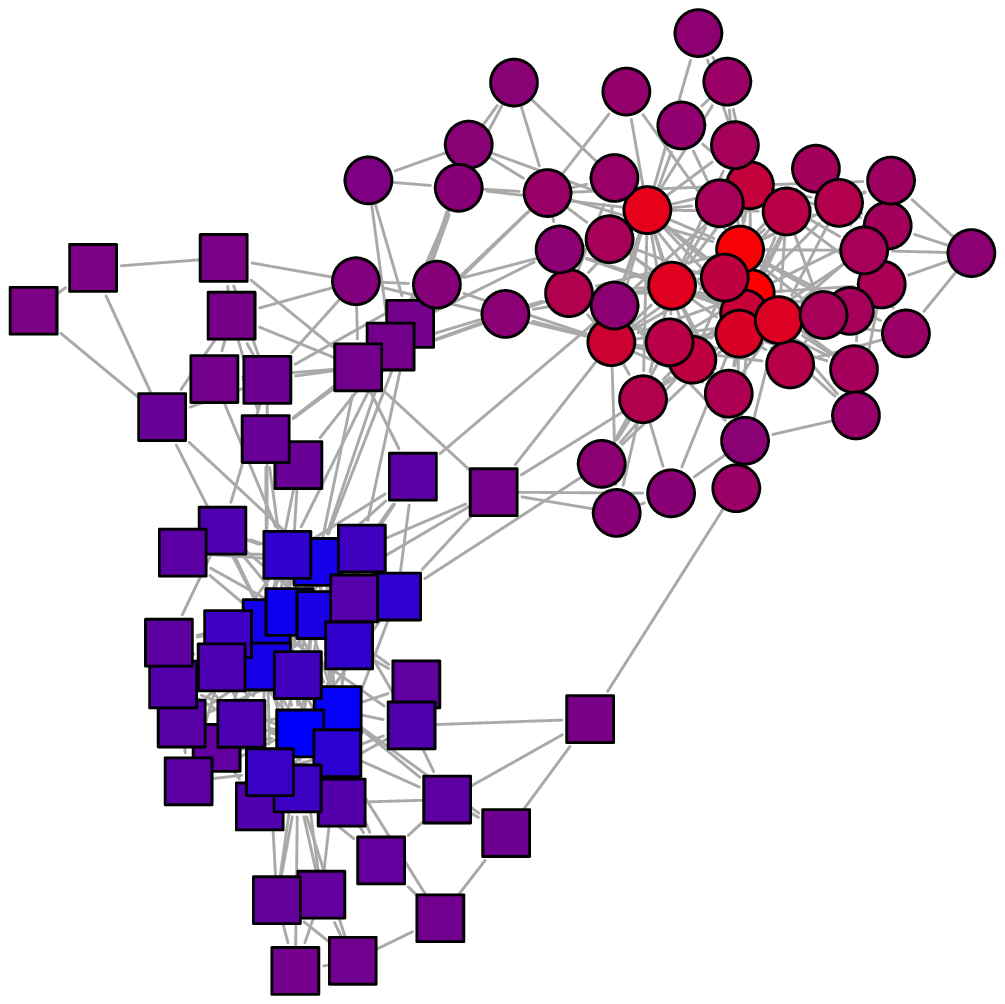}{5cm}{(a) Partition}{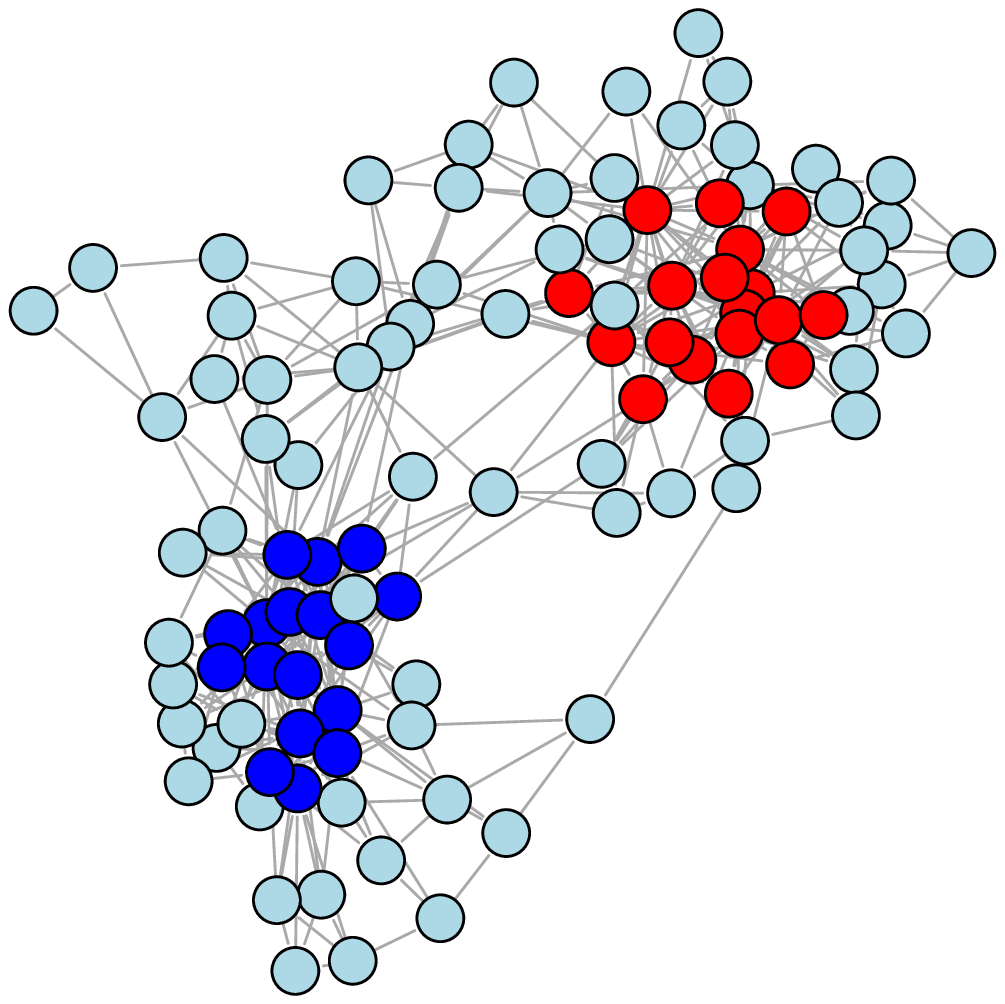}{5cm}{(b) Extraction}
\caption{Results for the political books network.  Node colors in (a) represent the components of the principal eigenvector of the modularity matrix, and node shapes represent the partition.  }\label{fig:polbooks}
\end{center}
\end{figure}

The nodes in this network \cite{Newman2006} are 105 recent political books with links representing pairs of books reported by Amazon as ``frequently bought together''.  Following \cite{Newman2006},  we show the modularity solution with node colors representing the components of the leading principal vector of the modularity matrix in Figure \ref{fig:polbooks}(a).  These values result from relaxing the labels from $\pm 1$ to real-valued, and the modularity partition is computed from the signs of these values (represented by node shapes).  The node colors can be interpreted to represent the book's position on the political spectrum, with blue being the most liberal and red the most conservative \cite{Newman2006}. Figure \ref{fig:polbooks} shows that in addition to a few clear ``red'' and ``blue'' books, many nodes are in 
fact ``purple'', and may not clearly belong to either the left or the right.  
From the colors alone (i.e., component magnitudes), it is not clear how to separate out the ``blue'' and the ``red'' from the more centrist ``purple'', whereas 
community extraction can do this easily.  Figure \ref{fig:polbooks}(b) shows the first two extracted communities which clearly 
correspond to the cores of the left and the right.  Further communities can be extracted, which we do not discuss here for lack of space. 

\subsection{The school friendship network}

\begin{figure}[h!]
\begin{center}
\centerline{\hfill\makebox[4.5cm]{(a) Grades}
\hfill\makebox[4.5cm]{(b) Partition with 6 groups}
\hfill\makebox[4.5cm]{(c) Extracting 6 groups}
\hfill\makebox[4.5cm]{(d) Extracting 7 groups}
            \hfill}
\centerline{
\hfill\includegraphics[width=4.5cm, height=5.5cm]{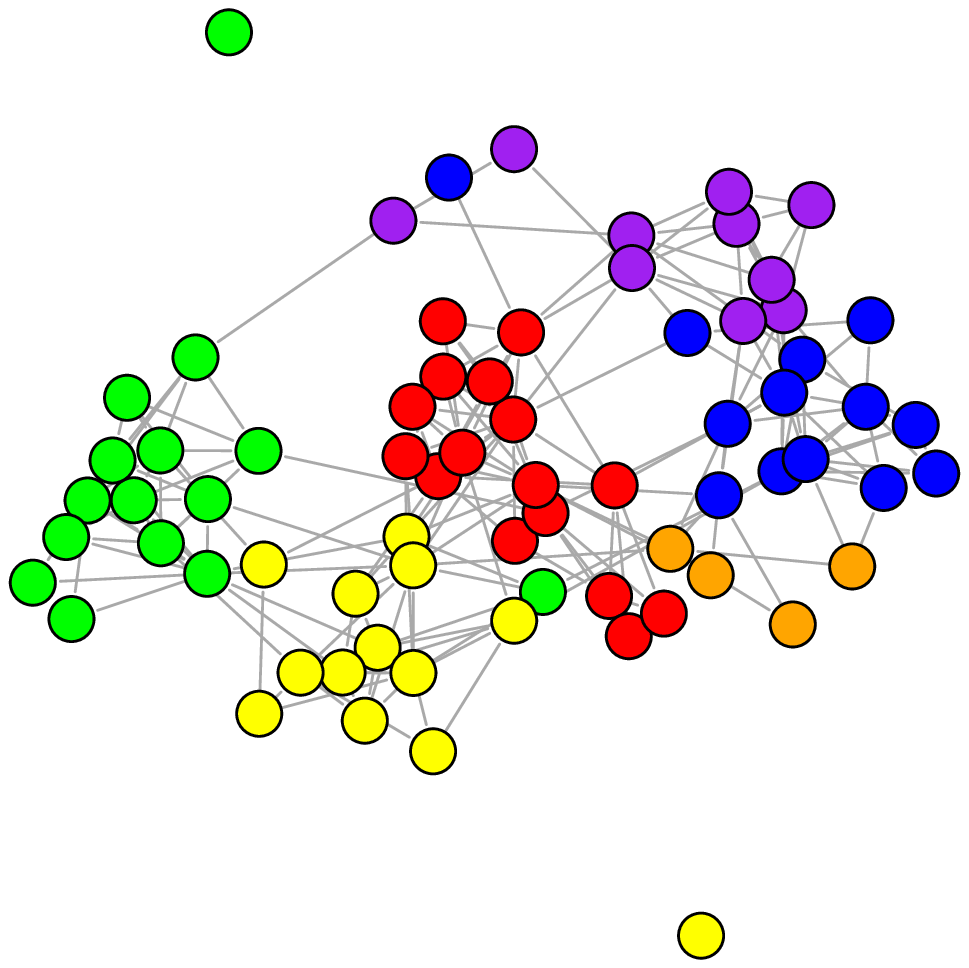}
\hfill\includegraphics[width=4.5cm, height=5.5cm]{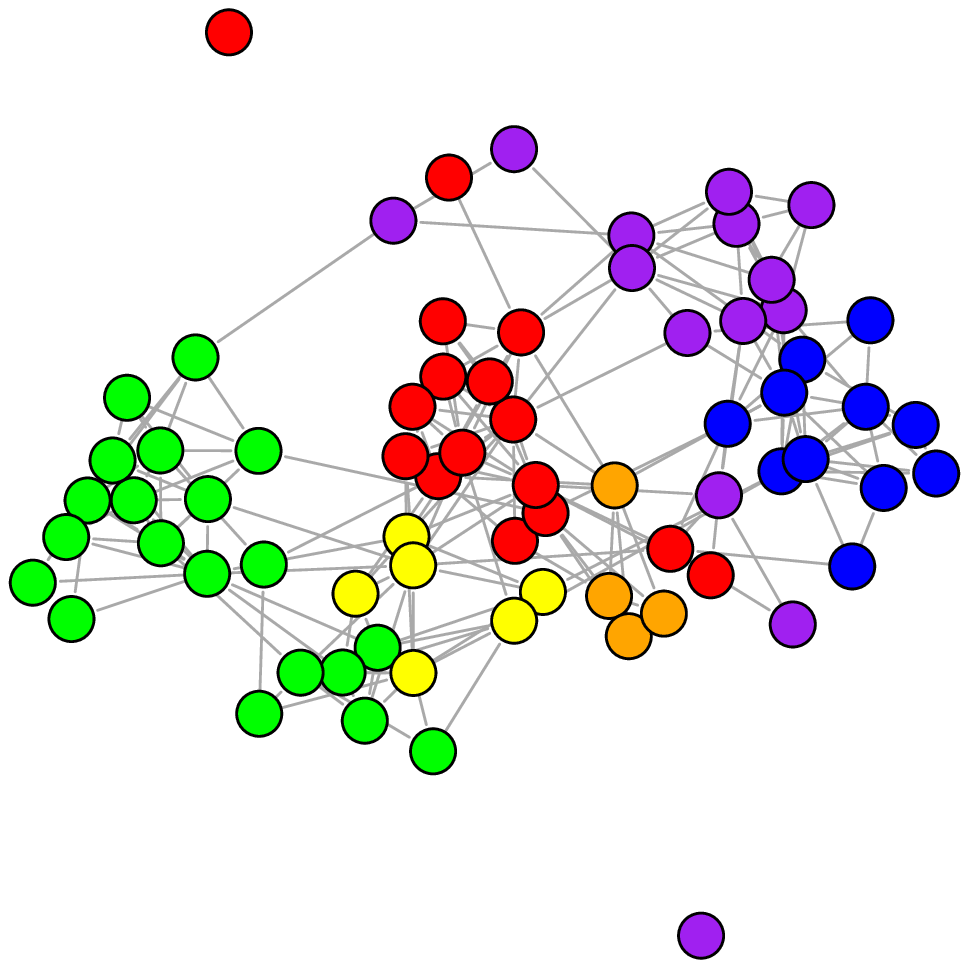}
\hfill\includegraphics[width=4.5cm, height=5.5cm]{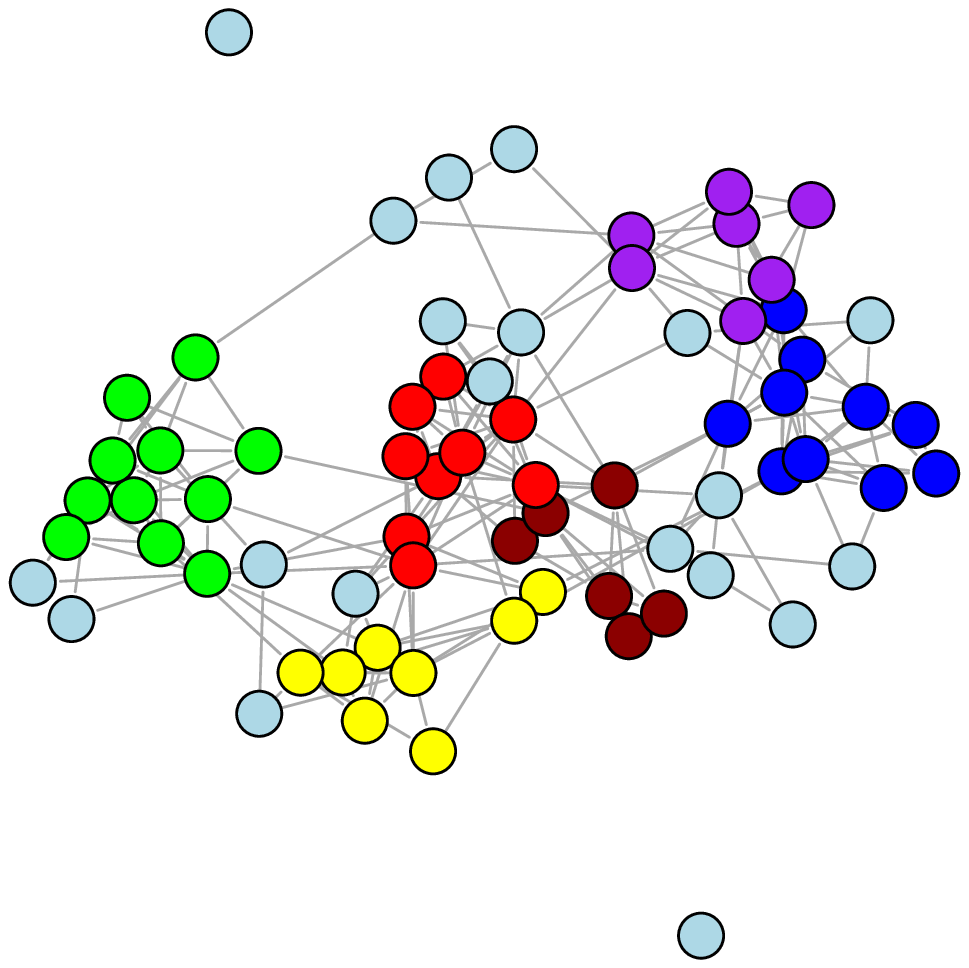}
\hfill\includegraphics[width=4.5cm, height=5.5cm]{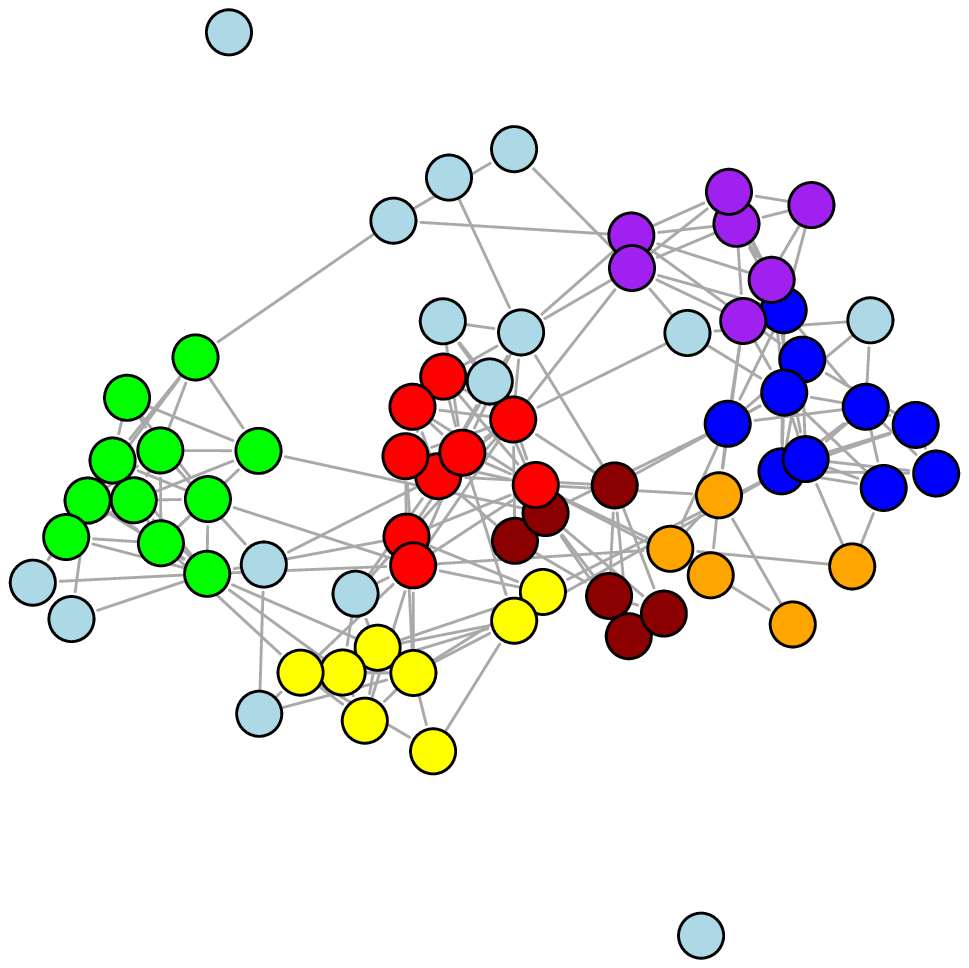}
           \hfill} 
\caption{Results for the school friendship network.}
\label{fig:school}
\end{center}
\end{figure}

This dataset is a school friendship network compiled from the National Longitudinal Study of Adolescent Health (see \cite{Hunter2008} for more information).   The survey asked students in grade 7 through 12 from 127 schools to name their close friends and answer a few questions to measure the strength of their friendships. Based on the answers, researchers 
constructed networks for each school with a weight 1-6 on each (directed) 
edge representing the strength of the friendship. Here we analyze the friendship network of 
school 1 from this dataset, converting the data to an undirected network by 
averaging the weights on the two edges connecting each pair of nodes. 
The resulting network with 71 nodes is shown in Figure \ref{fig:school} (a), with colors representing grades.  We show the results of extraction with six groups (to match the number of grades) in Figure \ref{fig:school}(c), and with seven groups, the number suggested by our stopping criterion, in Figure \ref{fig:school}(d). The seventh group picks up the grade shown in orange, which has only four nodes.  

For comparison, we partitioned the network into six communities by using 
the sequential procedure suggested in \cite{NewmanPNAS}, which partitions 
all communities from the previous step into two, and then the partition yielding the largest 
modularity is chosen to perform the next split. The modularity results shown in Figure \ref{fig:school}(b) are noticeably different from the grades themselves and from the extracted communities: a large part of the yellow grade is 
merged with green, and the smallest orange grade is split into three different groups.   This is partly a result of the greedy sequential splitting procedure, but also of the fact that modularity is forced to assign every single node to a community, with the extreme example being the two nodes that have no links at all.  The ability of our method to extract communities rather than partition into communities not only allows us to handle these and other weakly connected nodes correctly, but also appears to lead to more meaningful groups in this example.   

\section{Summary and Discussion}
We have proposed a new framework for analysis of social networks, which extracts tight communities out of the network and allows for background nodes.  In the examples we considered, it offers an additional insight into the network structure, and can be used as either an alternative or a complement to network partitioning.  While we have obtained good results with the tabu search, it may be beneficial to formulate the extraction criterion as an eigenvalue problem, which is work in progress.  More work is also needed on the stopping criterion and determining the correct number of communities;  this question, however, is common to all community detection methods.   Finally, assessing the quality of a proposed extraction/partition is an open problem;  one solution based on robustness to permutations was proposed in \cite{Karrer&Levina&Newman2008}, but assessing 
statistical significance under appropriate null models, and formulating such null models, is also of interest. 

\section*{Appendix: Proofs of Theorems}
Here we first state a simpler version of the main theorem of Bickel and Chen \cite{Bickel&Chen2009} 
sufficient for our purposes, and then apply it 
to prove Theorems 1 and 2.  The theorem holds for a general $K$ but to simplify notation we only state it for $K=2$.  We start from introducing additional notation.  
 Let $\M{O}$ be a $2\times 2$ matrix defined by 
$$O_{kl}(\V{s},\M{A})= \sum_{1\leq i,j \leq n} I(s_i=k, s_j=l) \ . $$
Evidently, $O_{kk}$ is twice the number of edges among nodes in the $k$-th community and $O_{kl}$ is the number of edges between the $k$-th and $l$-th 
communities. Let $\M{R}$ be 
the confusion matrix, 
$$R_{ab} (\V{s},\V{c})= \frac{1}{n} \sum_{i=1}^n I(s_i=a,c_i=b) \ , $$
where $\V{s}$ is a proposed label assignment and $\V{c}$ is the vector of true labels.  Finally, let $f(\V{c})=\M{R}^{T}\V{1}$ and $f(\V{s})=\M{R}\V{1}$ be  
the proportion of nodes in each block for the assignments $\V{c}$ and $\V{s}$,  respectively. 

Letting $\mu_n=n^2 \rho_n$, we can write the extraction criterion [{\bf 3}] (up to a multiplicative factor) in the form
$$Q(\V{s},\M{A})=F \left ( \frac{O(\V{s},\M{A})}{\mu_n},f(\V{s}) \right ) \ . $$Further, it is easy to verify that 
$$\frac{\mathbb{E} ( n^{-2} \M{O}(\V{s},\M{A})|\V{c})}{\mu_n} =\M{R}(\V{s},\V{c})\M{P}\M{R}^{T}(\V{s},\V{c}).$$
Thus the population version of $Q$ is $F(\M{RPR}^T,\M{R}\V{1})$.    
Then a natural necessary condition for asymptotic consistency of the criterion 
is that its population version is maximized by the 
correct diagonal confusion matrix, which gives us condition (C1) in the following Theorem by Bickel and Chen:

\begin{thma}
 Suppose $F$, $\M{P}$ and $\pi$ satisfy the following conditions: 
\begin{itemize}
\item [(C1)] $F(\M{RPR}^T,\M{R}\V{1})$ is uniquely maximized over $\mathscr{R}=\{\M{R}:\M{R}\geq 0, \M{R}^T\V{1}=(\pi,1-\pi)' \}$ by $\M{R}=\M{D}(\pi)\equiv diag(\pi,1-\pi)$, for all $(\pi,\M{P})$ in an open set $\Theta$.

\item [(C2)] $\M{P}$ has no identical columns.

\item [(C3)] (a) $F$ is Lipschitz in its arguments; 
(b) Let $\M{W}=\M{D}(\pi)\M{P}\M{D}(\pi)$.  The directional derivatives $\frac{\partial^2 F}{\partial \epsilon^2}(\M{M}_0+\epsilon(\M{M}_1-\M{M}_0), \V{t}_0+\epsilon(\V{t}_1-\V{t}_0))|_{\epsilon=0+}$ are continuous in $(\M{M}_1,\V{t}_1)$ for all $(\M{M}_0,\V{t}_0)$ in a neighborhood of $(\M{W},\V{C}(\pi))$, where $\V{C}(\pi)=(\pi, 1-\pi)^T$; 
(c) Let $G(\M{R},\M{P})=F(\M{RPR}^{T},\M{R}\V{1})$.  Then on $\mathscr{R}$, $\frac{\partial G((1-\epsilon)\M{D}(\pi)+\epsilon \M{R},\M{P})}{\partial \epsilon}|_{\epsilon=0+}<-C<0$ for all $(\pi,\M{P})\in \Theta$.
\end{itemize}
If $\hat{\V{c}}^{(n)}$ is the maximizer of $Q(\V{s},\M{A})$ and $\frac{\lambda_n}{\log n} \rightarrow \infty$, then, for all $(\pi, \M{P})\in \Theta$,
$$\limsup_{n\rightarrow \infty} \frac{P(\hat{\V{c}}^{(n)}\neq \V{c})}{\lambda_n}\leq -s_Q (\pi, \M{P})<0.$$
\end{thma}

\textbf{Proof of Theorem 1.}
Condition (C2) holds trivially and it is straightforward 
to check condition (C3), so we only check the essential condition (C1). We have
$W(S)  = \rho_n F (\M{O} / \mu_n,f(\V{s}))$,
where  $F(\M{M},\V{t})  = M_{11} / t_1^2 - M_{12} / (t_1t_2)$.
Thus, the population version of the criterion can be written as 
\begin{align*}
Q(\M{R}, \M{P}) = & \frac{1}{(r_{11}+r_{12})^2}(r_{11}^2 p_{11}+2r_{11}r_{12}p_{12}+r_{12}^2p_{22}) \\
        &-\frac{1}{(r_{11}+r_{12})(r_{21}+r_{22})}(r_{11}r_{21}p_{11}+ r_{11}r_{22}p_{12}+ r_{12}r_{21}p_{12}+ r_{12}r_{22}p_{22}) \ .
\end{align*}
We need to maximize this function over $\M{R}$ under the constraint 
$\M{R}^T \M{1} = (\pi, 1-\pi)^T$. 
Taking the transformation $t_1= r_{11} / (r_{11}+r_{12})$, 
$t_2= r_{22} / (r_{21}+r_{22})$,
we obtain
\begin{align*}
f= & p_{22}-p_{12}+(p_{11}-2p_{12}+p_{22}) \left [t_1(t_1+t_2-1)-\frac{1}{2}(t_1+t_2) \right ]+\frac{1}{2}(p_{11}-p_{22})(t_1+t_2).
\end{align*}
It is easy to verify that the function 
$g(t_1, t_2)=t_1(t_1+t_2-1)- (t_1+t_2)/2$
has two maximizers, $t_1=1,t_2=1$ and $t_1=0,t_2=0$. Thus under the condition $p_{11}-2p_{12}+p_{22}>0, p_{11}>p_{22}$, the unique maximizer of $f$ is 
$t_1=1,t_2=1$, or equivalently, $\M{R} = \F{diag}(\pi,1-\pi)$.   \hfill $\Box$

\textbf{Proof of Theorem 2.}
Again, we only verify (C1), since (C2) and (C3) are straightforward. For the adjusted criterion, we need to maximize 
\begin{align*}
f= & (r_{11}+r_{12})(r_{21}+r_{22}) \left [\frac{1}{(r_{11}+r_{12})^2}(r_{11}^2 p_{11}+2r_{11}r_{12}p_{12}+r_{12}^2p_{22}) \right .\\
        & \left. -\frac{1}{(r_{11}+r_{12})(r_{21}+r_{22})}(r_{11}r_{21}p_{11}+ r_{11}r_{22}p_{12}+ r_{12}r_{21}p_{12}+ r_{12}r_{22}p_{22}) \right ]\\
\end{align*} 
under the constraint $\M{R}^T \M{1} = (\pi, 1-\pi)^T$. Applying the same 
transformation $t_1= r_{11} / (r_{11}+r_{12})$, $t_2= r_{22} / (r_{21}+r_{22})$, we obtain
\begin{align*}
f= & \frac{(t_1-\pi)(t_2-(1-\pi))}{(t_1+t_2-1)^2}\left \{p_{22}-p_{12}+(p_{11}-2p_{12}+p_{22}) \left [t_1(t_1+t_2-1)-\frac{1}{2}(t_1+t_2) \right ] \right. \\
   & +\left. \frac{1}{2}(p_{11}-p_{22})(t_1+t_2) \right \},
\end{align*}
where $(t_1,t_2) \in [0,\pi]\times [0,1-\pi] \; \cup \; [\pi,1]\times [1-\pi,1]$.
The only interior point $t^*$ which potentially satisfies $\nabla f(t^*)=0$ is 
$$t^*_1=\frac{p_{22}-p_{12}}{p_{11}+p_{22}-2p_{12}}\quad \quad t^*_2=\frac{p_{11}-p_{12}}{p_{11}+p_{22}-2p_{12}}. $$ 
However, since $t^*_1+t^*_2=1$, the only intersection with the feasible region is at $t^*_1 =\pi$, $t^*_2=1-\pi$, and thus $f$ can only be maximized on the boundary of the feasible region.  Since all functions involved are monotone and convex, it is easy to check the boundary values; 
comparing all possible solutions shows that the unique maximizer of $f$ is 
$t_1=1,t_2=1$, or equivalently, $r_{11}=\pi,r_{22}=1-\pi$. \hfill $\Box$

\end{document}